\documentclass[twoside,11pt]{geophysicsf}

\usepackage{natbib}
\usepackage{amsmath}
\usepackage{graphicx}
\usepackage{color}
\usepackage{amssymb}
\usepackage{amsbsy}
\usepackage{lineno}
\usepackage{ulem}
\usepackage{setspace}
\usepackage{hyperref}
\usepackage{jmlr2e}

\setfigdir{Fig}

\def\beq{\begin{equation}}
\def\eeq{\end{equation}}
\def\beqa{\begin{eqnarray}}
\def\eeqa{\end{eqnarray}}

\begin{document}

\ShortHeadings{ML synthetic to real data}{Alkhalifah and Wang and Ovcharenko}

\author{\name Tariq Alkhalifah \email tariq.alkhalifah@kaust.edu.sa \\
\name Hanchen Wang \email hanchen.wang@kaust.edu.sa \\
\name Oleg Ovcharenko \email oleg.ovcharenko@kaust.edu.sa \\
\addr Physical Sciences and Engineering \\
King Abdullah University of Science and Technology \\
Thuwal 23955-6900 \\
Saudi Arabia}

\title{MLReal: Bridging the gap between training on synthetic data and real data applications in machine learning}


\maketitle

\begin{abstract}
Among the biggest challenges we face in utilizing neural networks trained on waveform data (i.e., seismic, electromagnetic, or ultrasound) is its application to real data. The requirement for accurate labels forces us to develop solutions using synthetic data, where labels are readily available. However, synthetic data often do not capture the reality of the field/real experiment, and we end up with poor performance of the trained neural network (NN) at the inference stage. We describe a novel approach to enhance supervised training on synthetic data with real data features (domain adaptation). Specifically, for tasks in which the absolute values of the vertical axis (time or depth) of the input data are not crucial, like classification, or can be corrected afterward, like velocity model building using a well-log, we suggest a series of linear operations on the input so the training and application data have similar distributions.  This is accomplished by applying two operations on the input data to the NN model: 1) The crosscorrelation of the input data (i.e., shot gather, seismic image, etc.) with a fixed reference trace from the same dataset. 2) The convolution of the resulting data with the mean (or a random sample) of the autocorrelated data from another domain. In the training stage, the input data are from the synthetic domain and the auto-correlated data are from the real domain, and random samples from real data are drawn at every training epoch. In the inference/application stage, the input data are from the real subset domain and the mean of the autocorrelated sections are from the synthetic data subset domain. Example applications on passive seismic data for microseismic event source location determination and active seismic data for predicting low frequencies are used to demonstrate the power of this approach in improving the applicability of trained models to real data.
  \end{abstract}

\newpage


\section{Introduction}

In our attempt to discover the Earth by inverting measurements of its physical properties we have developed hand-crafted algorithms to perform various tasks.
These hand-crafted (fixed) algorithms are independent of the specific features in the data, and usually, new research is required to update these algorithms to adapt to any evolution in the data to perform a specific task. Recently, data-driven approaches that have taken center stage in algorithms are replaced by dynamic neural network (NN) models
trained to perform specific tasks. 
Machine learning (ML) is gaining a lot of traction as a tool to help us solve outstanding problems in seismic waveform data processing and interpretation.
Most of the applications in our field have relied on supervised training of neural network (NN) models, where the labels (answers) are
available \cite[]{doi:10.1190/geo2017-0595.1,doi:10.1190/tle37010058.1,doi:10.1190/geo2018-0884.1,doi:10.1111/1365-2478.12865}. These answers are often available for synthetic data as we numerically control the experiment, or they are determined using human
interpretation or human crafted algorithms applied to real data.  The challenge in training our NN models on synthetic data is the generalization of the trained models
to real data, as that process requires careful identification of the training set and the inclusion of realistic noise and other variables between synthetic
and real data. In other words, the synthetic and real data are usually far from being drawn from the same distribution, which is essential for the success of a trained NN model \cite[]{Kouw2018AnIT}.
Thus, many synthetically trained NN models have performed poorly on real data. On the other hand, training on real data provides models that are often, at best, as good
as the accuracy of the labels that were determined by humans or human-crafted algorithms (weak supervision). So the data-driven feature of machine learning, in this case,
will be highly weakened \cite[]{10.1093/nsr/nwx106}.

Modeling is a general term encompassing any forward operation in which we have a model and we seek to obtain the corresponding data. The process can be as easy as a matrix-vector multiplication (linear) that can be used, for example,
to mute potentially missing traces from common shot gathers or image data for interpolation objectives, where the shot gathers or the images represent the model. Modeling can also include more complex operations like solving the wave equation using any numerical method \cite[]{aki2002quantitative}.
For seismic inversion, modeling, in its simplest form, is often given by the convolution of the source wavelet with a reflectivity model \cite[]{wang2016seismic}. Modeling is a deterministic process, and thus, the input-output combination is often unique and can be determined numerically. As such, using modeling and simulation, we can generate ML training data with their corresponding labels in a straightforward manner. Usually, in this case, the synthetically generated data (like shot gathers with missing traces) represent the input to the NN network in an attempt to have it learn to output the model (which is correspondingly the full shot gather). For the trained NN models on such synthetic data to generalize to real (application) data, the synthetic training set has to include the features embedded in the real data as much as possible \cite[]{Kouw2018AnIT}. For one, the training dataset (inputs and labels) should be represented by distributions that include the input and expected labels for the real data. However, this requirement, especially with respect to the input data to the network, is hard to achieve considering the simplified assumptions we use in modeling and simulation. It requires that we accurately reproduce the correlated and uncorrelated noise ($n(t)$) present in the real data, where $t$ here is the time (or any vertical axis like depth). More importantly, it also requires that we properly represent the source wavelet ($s(t)$) and the reflectivity ($r(t)$), present in the real data. In other words, for the synthetic data $d_s=r_s(t) \ast s_s(t) + n_s(t)$ in our simplified Earth model, the distributions of these three functions should cover those of the real data, and this is very hard to accomplish \cite[]{wreo20542}. Here $\ast$
stands for the convolution process.

The concept of trying to bridge the gap between the training and application data in machine learning is referred to as domain adaptation \cite[]{Kouw2018AnIT,Lemberger2020APO}. In this case, the training dataset is assumed to belong to the source domain and the application/testing data are assumed to belong to the target domain, the target of our training. The classic
theory of machine learning assumes that the application (target) data of a trained model should come from the same general population (sampled from the same distribution) as the training (source) set. So we need the probability distribution of the synthetic (source) dataset, $P_s({\bf x_s},{\bf y_s})$, where ${\bf x_s}$ are the inputs (i.e. synthetic waveform data), and ${\bf y_s}$ are the labels
(i.e. traveltime picks or horizons) for the source set, to equal the probability distribution of the real (target) dataset, $P_t({\bf x_t},{\bf y_t})$, where ${\bf x_t}$ are the inputs (i.e. real seismic data), and ${\bf y_s}$ are the labels for the target set that we often seek to predict.
One category of data adaptation is referred to as subspace mapping (or more generally, alignment) in which we find a transformation, $T$, that results in the distribution of the training (source) input data to equal that of the application (target) input data \cite[]{Fernando2013UnsupervisedVD}.
Specifically, $P_s \left(T({\bf x})\right)=P_r({\bf x})$. This can be accomplished by projecting the source and target data to the eigenvectors of the two subspaces, then finding a transformation between these projected spaces. Such projections can be achieved by Neural Network embedding aimed to find the weights of the embedding that minimizes the distance between the distribution of the source samples and the target ones.
There are many ways to constrain the transformation or weights to make the distributions similar including
the use of optimal transport \cite[]{villani2008optimal}. In this category, even cycle Generative Adversarial Networks (GANs) are used for the purpose of learning a 
generator to map target data to source data \cite[]{Gupta2019CycleGANFS}. 
However, these methods become more difficult to apply when the dimensions of the data are large, as is the case
with waveform (including seismic and ultrasound) data.  The method proposed in this paper shares the general framework of subspace alignment implemented in an empirical fashion.
 
 For seismic applications, as well as many other applications, we often acquire waveform data with sensors placed on the surface of the investigated
 body. In imaging applications, such data often loosely constrain the vertical (depth) axis (mostly described by the behavior of features along recording channels), and thus, we face 
 issues in imaging events accurately in depth. In this case, wells are often used to correct the depth misties as wells are considered ground truth markers \cite[]{doi:10.1190/segam2019-3211214.1}. Nevertheless, the structural information in these images is provided by the seismic (waveform) data. In applications, not requiring precise vertical dimension labeling (scale-wise), 
 crosscorrelation of the input data with a reference trace from the same data has recently been suggested to help reduce the variance in the distributions of the training and testing data, which ultimately helped with the application of ML for direct waveform microseismic event location \cite[]{doi:10.1190/geo2020-0636.1}. Prior to that, \cite{doi:10.1190/geo2010-0210.1} used the process of convolving a reference trace from the observed data with the synthetic data, and conversely convolving a reference trace from the synthetic data with the observed data in a waveform optimization problem to invert for the velocity model. This process helped reduce the difference between the terms of the objective function, and especially in mitigating the waveform source effect. In the sprite of these two processes, we propose a combination that could help us adapt the NN model training on synthetic data to work on real data.

An objective of a neural network model is to provide us with an output for a given input. The output that a trained model will give is based on the
training it experienced, and that depends mostly on the source training set and its distribution. 
A trained neural network model generalizes well when the target data are represented, as much as possible, in the source data set. To help
accomplish that when the application (target) data are real (field) data, we propose, here, to inject as much of the real data features into the synthetic data training as possible.
This can be accomplished by utilizing a combination of linear operations including crosscorrelation, autocorrelation, and convolution between the synthetic
and field data. These operations will bring the distributions of the training (source) synthetic dataset, and the (target) real dataset closer to
each other, which will help the trained model generalize better on real data. One real data generalization example we use here to test the approach is an NN model dedicated to locating microseismic sources directly from recorded waveform data. We also test the approach on an NN model trained to predict low-frequency data from high-frequency data to ultimately help full waveform inversion (FWI) converge better.

\section{An empirical Data projection}

The method proposed here is applicable mainly to supervised learning. Also, we assume that the vertical axis of the input sections (images or shot gathers) are not crucial in absolute values to the task at hand. Only the relative relation between events matters along that dimension. The reason for this assumption will be clear later.

\subsection{Setup}

We assume we do not have labels for the application data, and thus, we cannot perform transfer learning (a form of domain adaptation). In our case, the source synthetic data are labeled, but the target real data are not.  This form of domain adaptation is often addressed with unsupervised ML methods. 
In such domain adaptation, another important assumption is implied, and that is the target labels are drawn from the same distribution as the source labels ($P({\bf y_s})=P({\bf y_t})$). This is an
important assumption for the application of synthetically trained NN models on real data.  Of course, this requires that the task we plan to perform on the field data is represented
by the synthetic training data. Specifically, $P_s({\bf y_s}|{\bf x_s})=P_t({\bf y_t}|{\bf x_t})$. In physical terms, this implies that the modeling we do for our synthetic data generation represents the actual physical behavior. In other words, the assumptions used to model the synthetic training set represent the object of application (like the Earth). Thus, the issue we are addressing here is the case when the input distributions for the source and target data are not the same, specifically
$P_s({\bf x_s}) \neq P_t({\bf x_t})$.

This form of difference between the training and application datasets distributions in machine learning circles is referred to as a covariate shift \cite[]{Fernando2013UnsupervisedVD}. There are many ways to measure such
a shift, including using the Kullback-Lebler (KL) divergence metric. In domain adaptation, and similar to error bounds defined for machine learning in general, we can define an error bound on the 
application of a trained network model. This error bound is given by the combination of the error in the training and a term related to the complexity of the NN model (like its size). This, however, assumes that the training and application data come from the same distribution. For the case of a covariate shift, we have a similar, more complicated bound, given by \cite[]{NIPS2006_b1b0432c,Lemberger2020APO}:
\beq
\varepsilon_t(\mathcal{NN}) \leq \varepsilon_s(\mathcal{NN}) + d(P_s({\bf x_s}),P_t({\bf x_t})) + \lambda,
\label{eq:eq22}
\eeq
where $\varepsilon_s$ is the bound on the training error, and $d(.,.)$ is the distance between the marginal distributions of source and target datasets.
Here, $\lambda$ represents the optimal joint errors of a neural network model between the source and target datasets. So the upper bound of the application error
is guided by these three terms. 

So our objective is to devise a transformation that minimizes the difference measure $d$ in equation~\ref{eq:eq22} between the distribution of the training (source) and testing/application (target) datasets.
Specifically, we aim to find the transformation ${\bf \hat{x}_s}=T_s({\bf x_s})$ on the source data set and ${\bf \hat{x}_t}=T_t({\bf x_t})$ on the target data set so that the probability distributions $P_s({\bf \hat{x}_s}) \approx P_t({\bf \hat{x}_t})$.
Figure~\ref{fig:diagramD} summarizes this goal within the framework of the training process. So the input for the training of the Neural network ($\mathcal{NN}$) model
is ${\bf \hat{x}_s}$, in which the model parameters are optimized to match the labels ${\bf y_s}$ using a loss function ($\mathcal{L}$). On the other hand, the input during inference is ${\bf \hat{x}_t}$. We will discuss the transformations $T_s$ and $T_r$ in the next section, where the input are waveform (i.e. seismic) data $d$. 

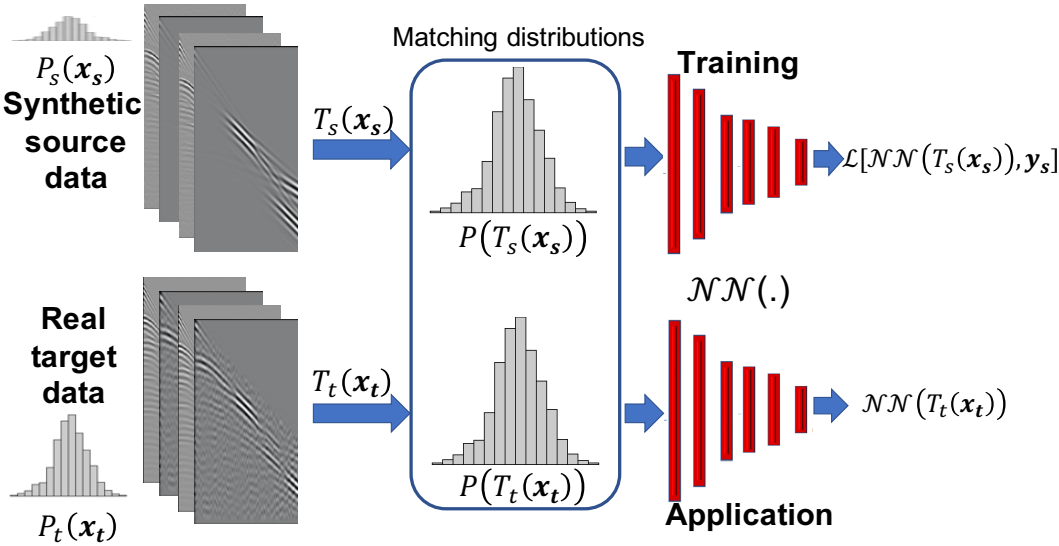
\plot{diagramD}{width=0.95\textwidth}{The workflow of the proposed data adaptation in which the training (source) dataset might have a different distribution than the 
 application (target) dataset, where transformations $T_s$ and $T_t$ will help reduce such differences and provide new data as input to the neural network 
function ($\mathcal{NN}$) to train the network to reduce loss $\mathcal{L}$, and then apply to real data. Here, $P$ are the probability distributions and we show
schematic versions of them for the source and target datasets.
}

\subsection{Conditioning synthetic data for training}

A trace in the seismic data can be represented by a combination of reflectively, source wavelet and noise, as follows:
\beq
d^{ij}(t) = r^{ij}(t) \ast s^{ij}(t) + n^{ij}(t),
\label{eq:eq1}
\eeq
where $i$ is the index of the trace, and $j$ is the index of the section in which the trace belongs to whether the section corresponds to a shot gather or a seismic image. Depending on the data, all three components $\left(r(t),s(t), n(t) \right)$ can vary over traces and sections. Here, $t$ may represent time or depth depending on whether the inputs are shot gathers or depth images, respectively. For a shot gather for example, often $r(t)$ changes with moveout, and of course, $n(t)$ changes from one trace to another. In training a neural network model to work properly on d(t), we often generate synthetic data, $d_s$
that hopefully includes a proper representation of these components. 

To do that, we use transformations applied on the vertical axis (time or depth), and thus, as mentioned earlier, we assume that the scale of the vertical axis is not crucial to the task. This assumption, though limiting the application of this approach, will allow us to apply the necessary transformation for domain adaptation. We will elaborate in the discussion section on the effect that this assumption may have on the 
application of this approach.  To migrate the components described in equation~\ref{eq:eq1} from the real data to the synthetic ones, we use linear operations, and thus, we define new training data (source data transformed by the proposed approach) as follows:
\beq
\hat{d^i_s}(t) = d_s^i(t) \otimes d_s^k(t) \ast d^{ij}(t) \otimes d^{ij}(t),
\label{eq:eq2}
\eeq
where $k$ is the index of a reference trace from the synthetic input data (fixed for both synthetic and real data), and $j$ is the index of a section from the real data whether the section corresponds to a shot gather or a seismic image. The operator $\otimes$ represents crosscorrelation,
and in this equation, we have a crosscorrelation between the input synthetic section $d_s^i(t)$ and a reference trace from that section $d_s^k(t)$ convolved with a randomly drawn ($j$) autocorrelated
section from the real data $ d^{ij}(t)$. The reference trace can be a near offset trace in the case of a shot gather input, or it can be any trace from the input for a seismic image.
The index of the reference trace should not change between sections to maintain the relative relation between sections. The randomly picked $j$ index for
autocorrelated real data varies in the training per epoch to allow for proper representation of the real data imprint on the training set.
Assuming we only have noise in the real data, the autocorrelation of random noise yields a quasi delta function at zero lag proportional to the energy of the noise. A convolution
with such a function will incorporate that energy into the synthetic data so that the signal-to-noise ratio (SNR) in the transformed synthetic data 
would be comparable to that of the autocorrelated real data. 

To allow for a transformed source data to have a similar distribution to the application data, we also transform the target (application) data in a similar fashion in which the features of the source synthetic data are incorporated in the target data.
To achieve that, we apply the following transformation, $T$, to the real data during the inference stage:
\beq
\hat{d^i}(t) = T(d^i) = d^i(t) \otimes d^k(t) \ast \frac{1}{N_s} \sum_j d_s^{ij}(t) \otimes d_s^{ij}(t),
\label{eq:eq3}
\eeq
which includes the same operations as in equation~\ref{eq:eq2} with the role of the real and synthetic data reversed. In this case, $N_s$ represents the number of
synthetic sections in the training set, and we actually convolve with the mean of the autocorrelated synthetic data, instead of a randomly drawn sample, like in the training. We will refer to the transformations in equations~\ref{eq:eq2} and~\ref{eq:eq3} as \{\textit{MLReal transformations}\}.
The idea of having two instances of each data (synthetic and real) in equations~\ref{eq:eq2} and~\ref{eq:eq3}
is to balance their contribution to the new training and testing data sets.  This way, we match the properties of the synthetic and real data used for training and inference, respectively.
Note that the convolution operation that connects equal amounts of the synthetic and real data
contributions is commutative. We will see the value of this operation more in the next section. 


Meanwhile, Figure~\ref{fig:traintestData2FF} (left)
demonstrates the process of applying equation~\ref{eq:eq2}, where the synthetic data were generated for the training of a NN model to predict low frequencies \cite[]{doi:10.1190/geo2018-0884.1}. The objective was to improve the full waveform inversion convergence for real marine data by adding low-frequency content into the data. On the other hand, Figure~\ref{fig:traintestData2FF} (right) demonstrates the process of applying equation~\ref{eq:eq3} on real data for an input to the trained model. Note that the resulting shot gathers look similar for the
two processes, and especially in the distribution of energy along the channels. They, however, still maintain the characteristics (the moveout) of the original data (synthetic or real). However, they contain more energy and more features in which the NN model can utilize. We will show the results of applying the proposed process for this task in the examples.

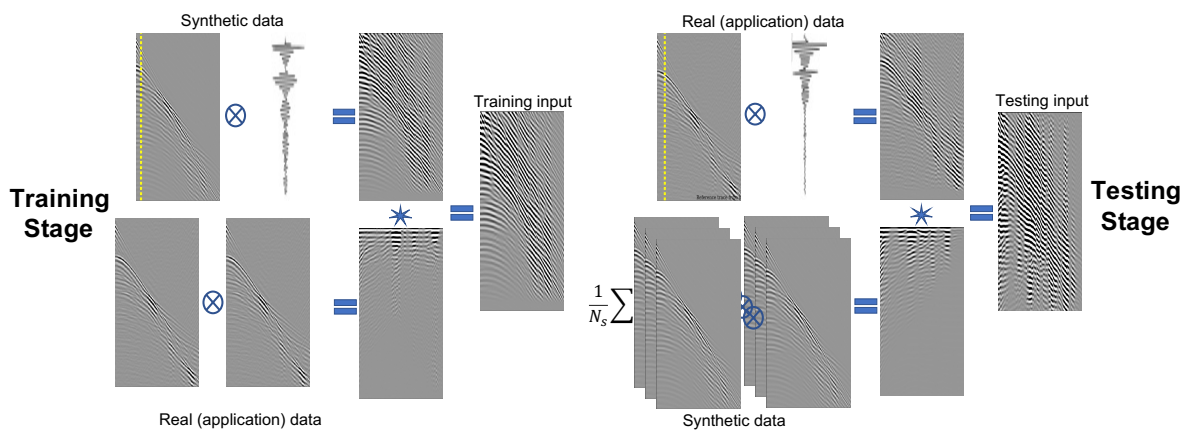
\plot{traintestData2FF}{width=1.05\textwidth}{The workflow chart for the MLReal method applied to marine data. Left side: The proposed process used for producing the training data; Right side: the proposed process used for producing the testing/application data. 
The circled cross symbol denotes a crosscorrelation operation, and the star symbol denotes a convolution operation. The dashed
vertical yellow line in the input traces indicates the location of the reference trace.}


\subsection{A frequency domain analysis}

If we transform the real data to the frequency (or vertical wavenumber) domain, considering the basic laws of the Fourier representation of crosscorrelation and convolution, equation~\ref{eq:eq1} can be written as
\beq
D^{ij}(\omega) = A^{ij}(\omega) e^{i \phi^{ij}(\omega)} = R^{ij}(\omega) S^{ij}(\omega) + N^{ij}(\omega),
\label{eq:eq4}
\eeq
where $\omega$ is the angular frequency, $A$ and $\phi$ are the amplitude and phase, respectively, of the complex-valued data. All capital letters represent the frequency domain form of the reflectivity, source, and noise functions in equation~\ref{eq:eq4}, given respectively. As a result, we can write equation~\ref{eq:eq2}
in the frequency domain as
\beqa
D^i_t(\omega) &=& \bar{D_s}^i(\omega) D_s^k(\omega) \bar{D}^{ij}(\omega) D^{ij}(\omega) = \bar{D_s}^i(\omega) D_s^k(\omega) \left(A^{ij}(\omega)\right)^2  \nonumber \\
   &=& \bar{D_s}^i(\omega) D_s^k(\omega) \left(R^{ij}(\omega) S^{ij}(\omega)+ N^{ij}(\omega)\right) \left(\bar{R}^{ij}(\omega) \bar{S}^{ij}(\omega)+ \bar{N}^{ij}(\omega)\right),
\label{eq:eq5}
\eeqa
where the overstrike, $\bar{.}$, symbol stands for the complex conjugate.

The application of the model on real data will involve an input to the model given by equation~\ref{eq:eq3}, which can be represented in the frequency domain by
\beqa
D^i_r(\omega) &=& \bar{D}^i(\omega) D^k(\omega)  \frac{1}{N_s} \sum_j \bar{D_s}^{ij}(\omega) D_s^{ij}(\omega) = A^{i}(\omega) A^{k}(\omega) e^{i \left(\phi^i(\omega) - \phi^k(\omega) \right)}  \frac{1}{N_s} \sum_j \bar{D_s}^{ij}(\omega) D_s^{ij}(\omega) \nonumber \\
                &=& \left(R^{i}(\omega) S^{i}(\omega)+ N^{i}(\omega) \right) \left(\bar{R}^{k}(\omega) \bar{S}^{k}(\omega)+ \bar{N}^{k}(\omega)\right) \frac{1}{N_s} \sum_j \bar{D_s}^{ij}(\omega) D_s^{ij}(\omega).
\label{eq:eq6}
\eeqa
Note that the frequency content of the input to the training and the application on real data will include a squared amplitude ($A$) of the real data in an even manner. Also, the key here is that equations~\ref{eq:eq5} and~\ref{eq:eq6} share a similar magnitude of noise, reflectivity, and source signature, or in more general terms, similar energy. Since the two data sets, after transformation,
share the same elements sampled from their corresponding distributions, then the distributions of the two data sets should be close, which allows us to satisfy the requirement
$P_s(T_s({\bf x})) \approx P_s(T_s({\bf r}))$ for the generalization of the NN model. In other words, the second term in equation~\ref{eq:eq22} will be small.

\section{A microseismic data example}

We will first show the impact of the above operations on real data acquired as part of monitoring microseismic events. The passive seismic acquisition 
was performed using a star configuration of sensors, as shown in Figure~\ref{fig:receivergeometry}, to monitor a hydraulic fracturing stimulation of a shale gas reservoir
in the Arkoma Basin in the United States. Figure~\ref{fig:raw_field_data} shows the real data for one microseismic event. The shot gather section includes ten segments from the various lines (azimuths), shown in Figure~\ref{fig:raw_field_data},
plotted here side by side. We were provided a total of 75 of these 
microseismic event recordings and the corresponding locations of the events were determined using conventional methods (we will use here only 10 of them). These labels (event locations) will help us evaluate the accuracy of our trained NN model. For more details on the data, we refer you to \cite{stanvek2017seismicity}. We were also given a velocity model for the area, 
which is shown in Figure~\ref{fig:velocitymodelvp}. Using this velocity model,  we employ a second-order in time and fourth-order in space, finite difference approximation to solve the acoustic wave equation and simulate wavefields from 5000 randomly placed seismic sources within the region of interest (the region we expect the real events to be located). The resulting 5000 synthetically recorded sections, using the layout in Figure~\ref{fig:receivergeometry}, and
the corresponding event location (labels) are split in a random manner into a training set (4000 samples) and a validation set (1000 samples). An example synthetic
data section, for an event near the one estimated for the real data section in Figure~\ref{fig:raw_field_data}, is shown in Figure~\ref{fig:raw_synthetic_data}. If we compare this synthetically generated section with the field one, we can appreciate the large difference between the two data despite them sharing similar general shapes as they originate from nearby sources. We do not have the source time information, which explains the shift between the events
in the two sections. We trained the neural network model on such synthetic data (used the crosscorrelation operation with a reference trace to mitigate the shift \cite[]{doi:10.1190/geo2020-0636.1}, and because of the large differences in the data between training and testing data, the accuracy of the location of the 10 events, as we will see later, is low.

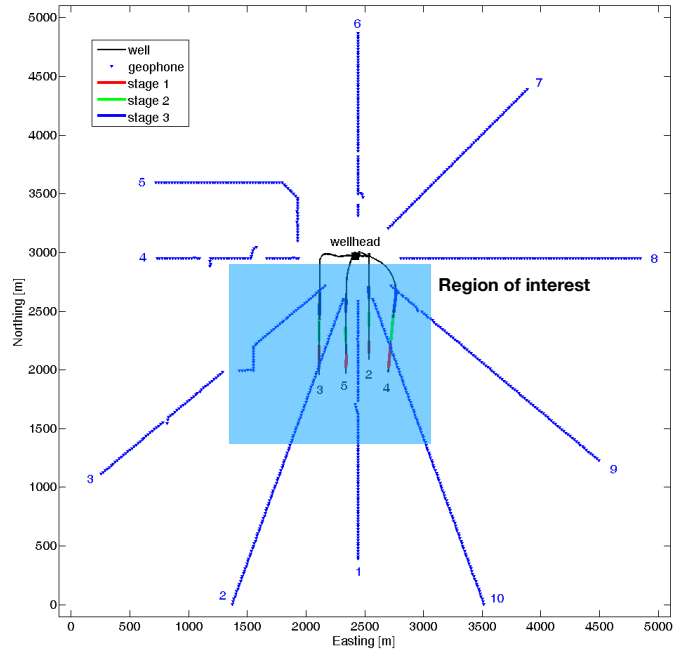
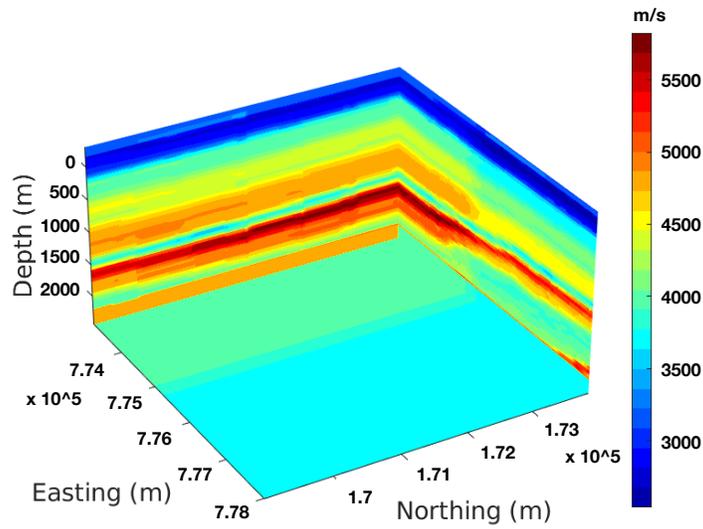
\multiplot{2}{receivergeometry,velocitymodelvp}{width=0.65\textwidth}{a) The passive seismic acquisition lines. b) The velocity model estimated in the region and used here to generate the synthetic data.}

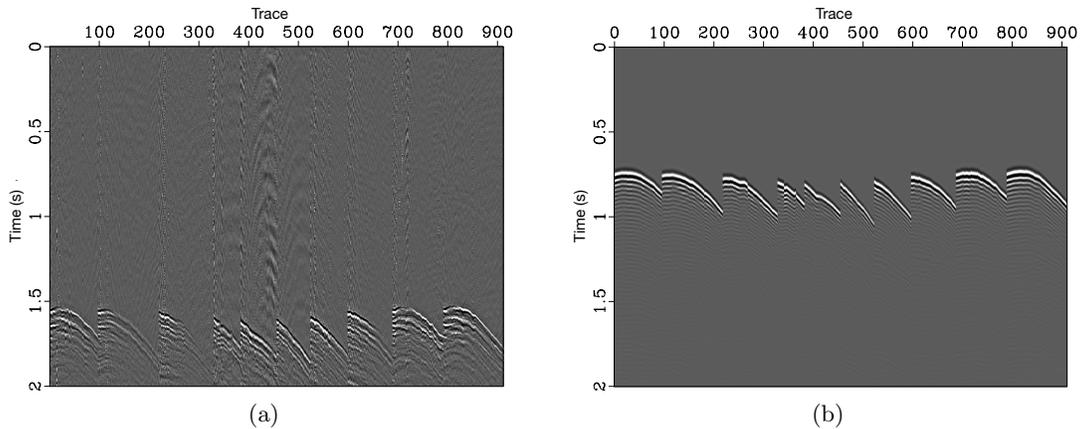
\multiplot{2}{raw_field_data,raw_synthetic_data}{width=0.45\textwidth}{a) The field recorded data for a single microseismic event along the 10 lines plotted side by side. b) The
synthetic data along the same lines from a source near the field data one, which was provided for this event. The time of the source is unknown, which explains the shift.}

\subsection{Data transformations}

We first calculate the crosscorrelation of the input section with a reference (fixed location for all data synthetic and real) trace from the section. The reference trace in this
application is given by the first trace of each input section, or in other words, the first trace of the first line in the section. An example result of this operation applied to the sections shown in Figures~\ref{fig:raw_field_data} and~\ref{fig:raw_synthetic_data},
is given by Figures~\ref{fig:field_data_correlate_with_reference_trace} and~\ref{fig:synthetic_data_correlate_with_reference_trace}, respectively. This operation has
largely mitigated the time shift between the field and synthetic data, as the vertical axes are now given by the time lag.
We then calculate the autocorrelation of the field and synthetic data, a sample of which are shown in Figures~\ref{fig:auto_correlation_field_data} and~\ref{fig:auto_correlation_synthetic_data} corresponding to the sections shown in Figures~\ref{fig:raw_field_data} and~\ref{fig:raw_synthetic_data}, respectively. We can appreciate how much
the autocorrelation of the field section carries information representing the source wavelet and the noise, in a zero-phase fashion. 
Applying the operations involved in the MLReal transformations (equations~\ref{eq:eq2} and~\ref{eq:eq3}) on the synthetic and real data, we obtain the sections shown in Figures~\ref{fig:field_data_convolve_with_mean_autocorrelation2}
and~\ref{fig:synthetic_data_convolve_with_mean_autocorrelation2}, respectively.
We window the part around zero lag to reduce the size of the input-to-the-network data.
The two sections look much more alike than those in Figures~\ref{fig:raw_field_data} and~\ref{fig:raw_synthetic_data}. 

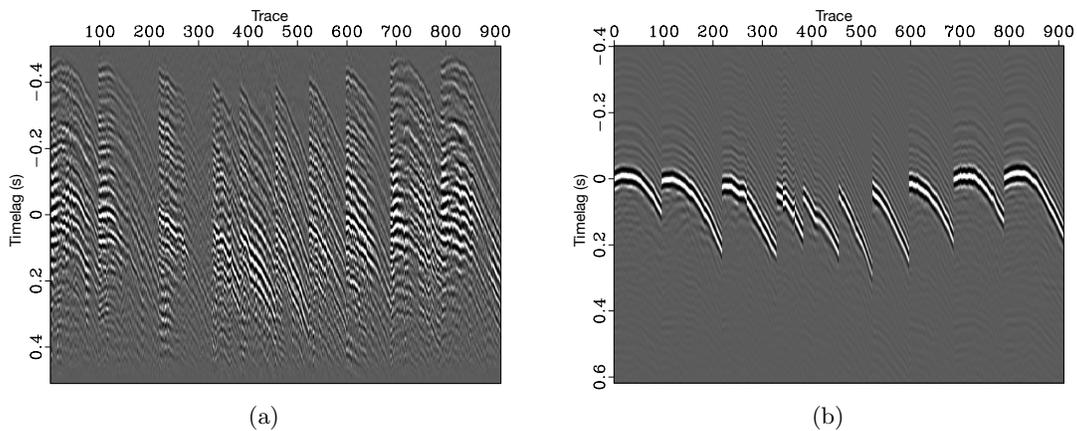
\multiplot{2}{field_data_correlate_with_reference_trace,synthetic_data_correlate_with_reference_trace}{width=0.45\textwidth}{a) The crosscorrelation of the field recorded data shown in Figure~\ref{fig:raw_field_data} with the first trace in the section. b) The crosscorrelation of the synthetic data shown in Figure~\ref{fig:raw_synthetic_data}
with the first trace in the section.}

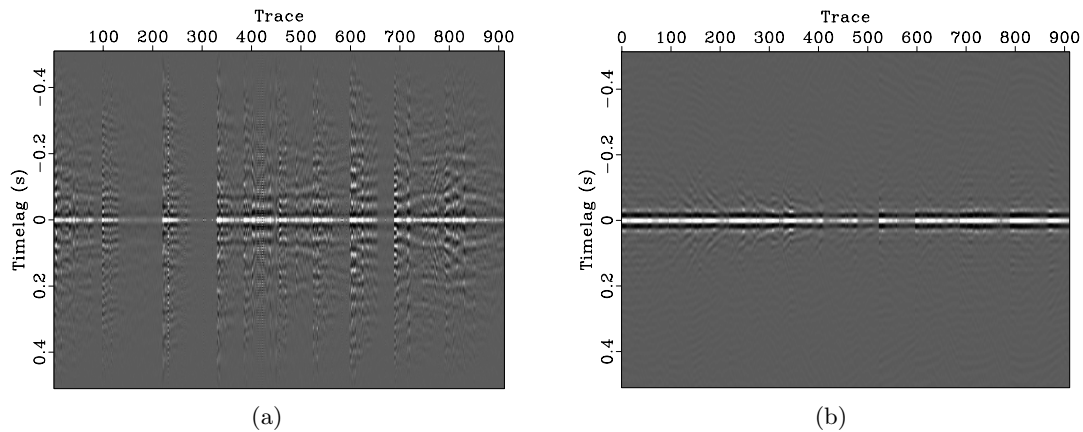
\multiplot{2}{auto_correlation_field_data,auto_correlation_synthetic_data}{width=0.45\textwidth}{a) The autocorrelation of the field recorded data shown in Figure~\ref{fig:raw_field_data}. b) The autocorrelation of the synthetic data shown in Figure~\ref{fig:raw_synthetic_data}.}


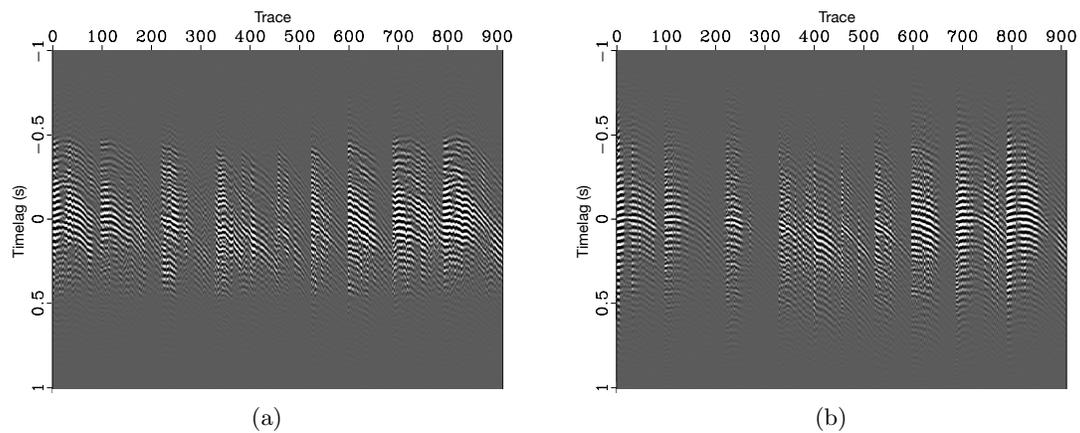
\multiplot{2}{field_data_convolve_with_mean_autocorrelation2,synthetic_data_convolve_with_mean_autocorrelation2}{width=0.45\textwidth}{a) An example section
 testing application data after applying equation~\ref{eq:eq3} (the proposed transformation) for the section shown in Figure~\ref{fig:raw_field_data}.
 b) The training input data corresponding to the section shown in Figure~\ref{fig:raw_synthetic_data} after injecting it with real data information using the proposed method, and specifically equation~\ref{eq:eq2}.}

\subsection{The training}

Using sections like that shown in Figure~\ref{fig:synthetic_data_convolve_with_mean_autocorrelation2} in which the location of the source (as the label) is known from modeling,
we train a 14-layer convolutional neural network to predict the location of the microseismic source ($x$, $y$, and $z$). The training included 4000 input sections (samples, modeled synthetically) and their corresponding labels (the training set),
and the training was executed over 5000 epochs, single batch, using an Adam optimizer \cite[]{kingma2014adam}. The convolution with randomly selected auto correlated sections from the real data is done at every one of the 5000 epochs. The random selection allows for more variance in the information extracted from the real data to be present
in the training samples. The loss function for the training and validation, shown in Figures~\ref{fig:loss_training} and~\ref{fig:loss_validation}, respectively, show good convergence.
As we may expect due to the data being recorded at the surface, the error in the horizontal location of an event is far less than that for the vertical location. Thus,
the lateral resolution in locating the event is expected to be higher.

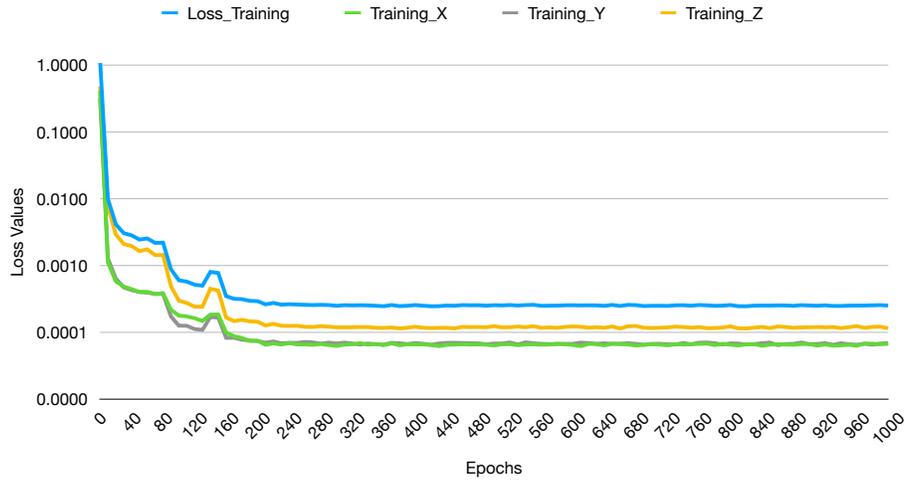
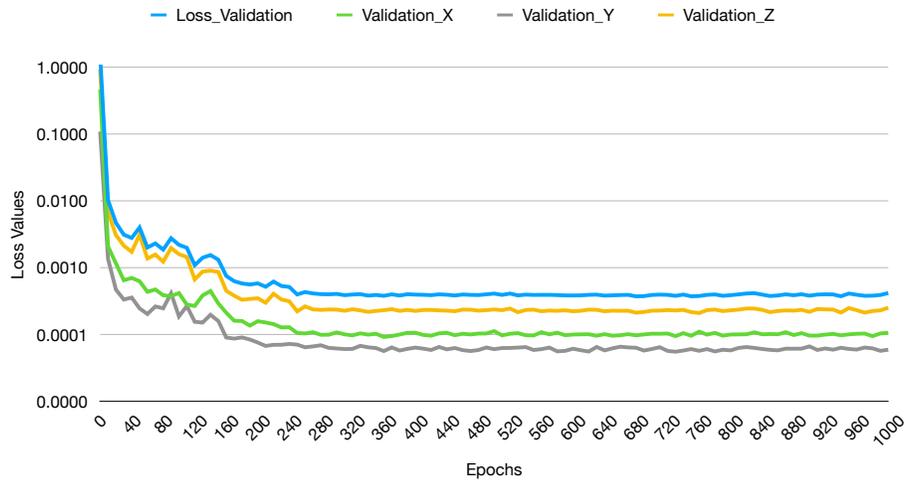
\multiplot{2}{loss_training,loss_validation}{width=0.8\textwidth}{a) The training loss for the  separate coordinate components  as well as the total loss (distance). b) The validation losses.}

\subsection{Results}

Then we input 10 real data sections, after applying the operations in equation~\ref{eq:eq3}, into the trained model to evaluate the accuracy
of the prediction. Figure~\ref{fig:event_location_3D_given_vs_autocor_vs_no_autocor_new} shows the predicted locations (in blue) and the provided ones (in red) in the region of investigation. We also show the predicted without convolving with the autocorrelation, just the crosscorrelation with a reference trace \cite[]{doi:10.1190/geo2020-0636.1}.
The differences are generally small and can be caused by many factors. For one, the NN model is known to have a bias toward smoothing the output \cite[]{rahaman2019spectral}. A very small network \cite[]{doi:10.1190/segam2020-3425457.1}
will levitate towards the mean of the training labels.  Of course, a cure for that is to increase the network size.  Another reason for the difference could be the simplified assumptions
used in our data simulation for the training of our NN model compared to the modeling approach potentially used in determining the location of the events by the data providers.
To further evaluate the improvements in data coherency between training and application using the proposed MLReal transformations,
we compute the normalized Euclidean distance (NED) between the new testing (real) sections and the corresponding synthetic ones.
Figure~\ref{fig:LD_vs_NED} shows NED for sections in which only the crosscorrelation with a reference trace is used
to mitigate the shift (dashed blue), and those for sections in which MLReal transformations were used (solid blue) in the training. Though NED
values can range between 0 to 1, the values are reasonably higher for our proposed method, compared to just cross-correlating with a reference. We also plot in Figure~\ref{fig:LD_vs_NED}
the distances between the predicted sources and provided ones for the same 10 events. For our proposed method, these values average around 15 meters. 
Considering the quality of the
data and the layout of the sensors on the surface, the differences for the proposed method are very reasonable. Meanwhile, with only the crosscorrelation with a reference trace, 
which was developed earlier \cite[]{doi:10.1190/geo2020-0636.1},
the location difference averaged around 45 meters (Figure~\ref{fig:LD_vs_NED}, in black). So the transformations (our approach for domain adaptation) helped a lot in this example.

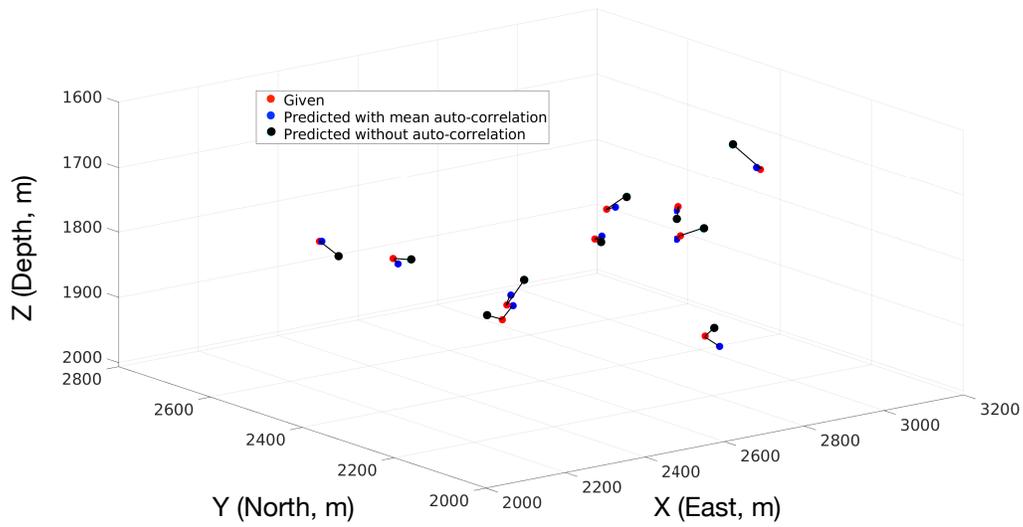
\plot{event_location_3D_given_vs_autocor_vs_no_autocor_new}{width=0.95\textwidth}{a) A 3D plot of the locations of the predicted (blue dots)
and provided (red dots) source locations for 10 field data events, as well as the trained and predicted without the convolution with the autocorrelation of the other data (grey dots).
The thin lines between the dots connect both predictions to the provided locations from both predictions.}

\multiplot{2}{event_location_XY_given_vs_autocor_vs_no_autocor_new,event_location_XZ_given_vs_autocor_vs_no_autocor_new,event_location_YZ_given_vs_autocor_vs_no_autocor_new}{width=0.45\textwidth}{The locations of the predicted (blue dots)
and provided (red dots) source locations for 10 field data events, as well as the trained and predicted without the convolution with the autocorrelation of the other data (black dots),
all of which are
projected on to a) the $x-y$ plane, b) the $x-z$ plane, and c)  the $y-z$ plane. The thin lines between the dots connects both predictions to the provided locations. }

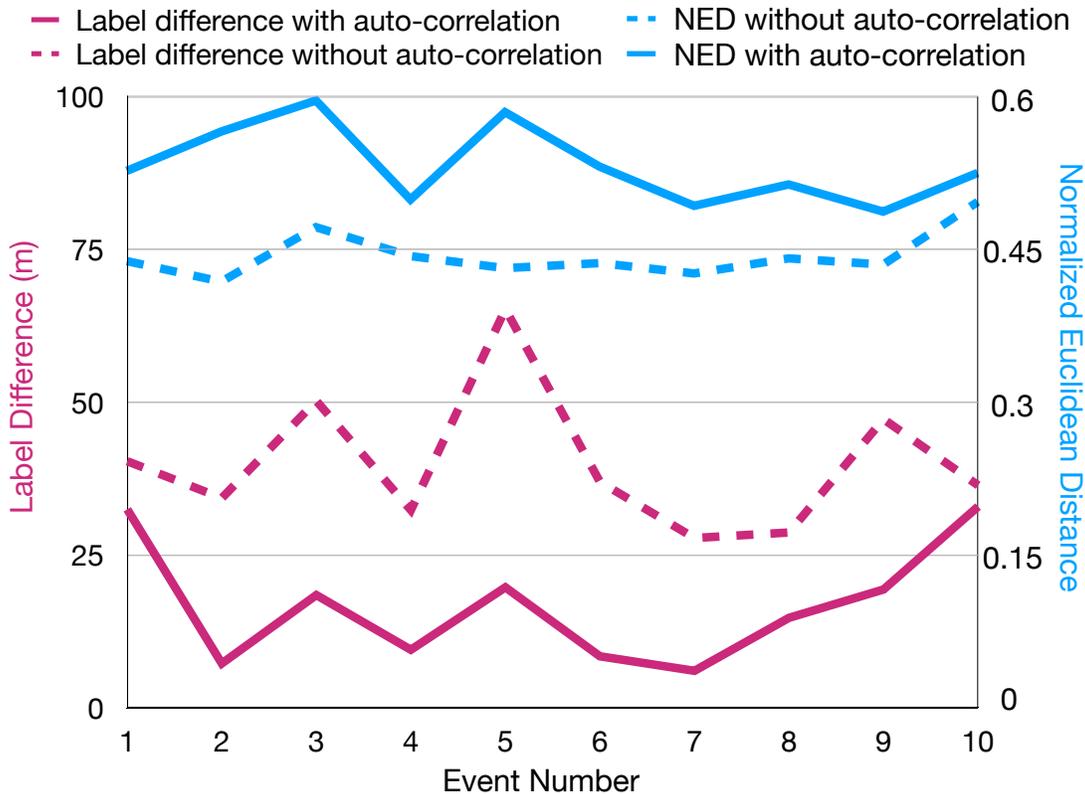
\plot{LD_vs_NED}{width=0.95\textwidth}{The normalized euclidean distance (blue) between the field data and modeled data
from the predicted source locations (blue), as well as the actual source location (the label) difference between the provided locations and the predicted locations using the proposed method (solid lines)
and using the same network trained without the convolution with the auto-correlation (dashed lines).}

\section{Low-frequency prediction example}

In this example, we apply the approach to the task of low-frequency extrapolation. In particular, we reconstruct low-wavelength components, $<$~5 Hz, for a complete shot gather from available high-frequency representation, $>$~4 Hz, of this shot gather. The predicted low-frequency data aim to help full-waveform inversion (FWI) converge to the global minimum by mitigating the cycle-skipping problem \cite[]{doi:10.1190/geo2018-0884.1}.
Since low frequencies are often not available in real-world field data or are contaminated with noise, we train a deep neural network on synthetic data and run inference on a band-limited marine dataset acquired offshore northwest Australia. An example shot gather is shown in Figure~\ref{fig:traintestData2FF}. 

Similar to an inverse problem, where the sought-after data are often unknown, available low-frequencies in the field dataset are insufficient to enable unsupervised training for bandwidth extrapolation. For this reason, we train the deep learning model on synthetic waveforms simulated in a variable range of elastic media initializations. Specifically, we extract the acquisition parameters and source signature from the field data and use these for numerical modeling in a set of synthetic subsurface initializations.
The source wavelet should have a broad spectrum to allow us to bandpass the modeled shot gather to a high-frequency shot gather (similar band to the field data) and a low-frequency band shot gather (our objective). Common shot gathers, generated using such a setup, provide inputs and labels for the training of the deep learning model. 

\subsection{Data transformations}

For training on synthetic data, we only apply the MLReal transformations on the inputs to the network assuming that we seek targets similar to those from the synthetic data distribution. In other words, we assume that the velocities used in generating the synthetic data are close to the true Earth one, which will mitigate the limitation we mentioned with regard to the vertical dimension of the input. To set up the adaptation workflow for the input high-frequency data, we select an arbitrary trace from the high-frequency partition of the synthetic dataset and use this trace as a constant amplitude reference throughout the application. During training, for a given sample of input high-frequency data, we first cross-correlate it trace-wise with the trace selected earlier (Figure~\ref{fig:traintestData2FF}). Then we draw a random high-frequency data sample from the field dataset and convolve its auto-correlation with the result of the previous operation, as demonstrated in Figure~\ref{fig:traintestData2FF}. 
The effect of such transformations is shown in Figure~\ref{fig:input_target}. The transformed synthetic and real data look-alike to the point they share similar energy distribution as a function of offset. The synthetic transformed data are used to train the network without the need to apply any alterations to the output synthetic low-frequency shot gather, as the NN needs to learn such mapping since the transformed real data have similar features.

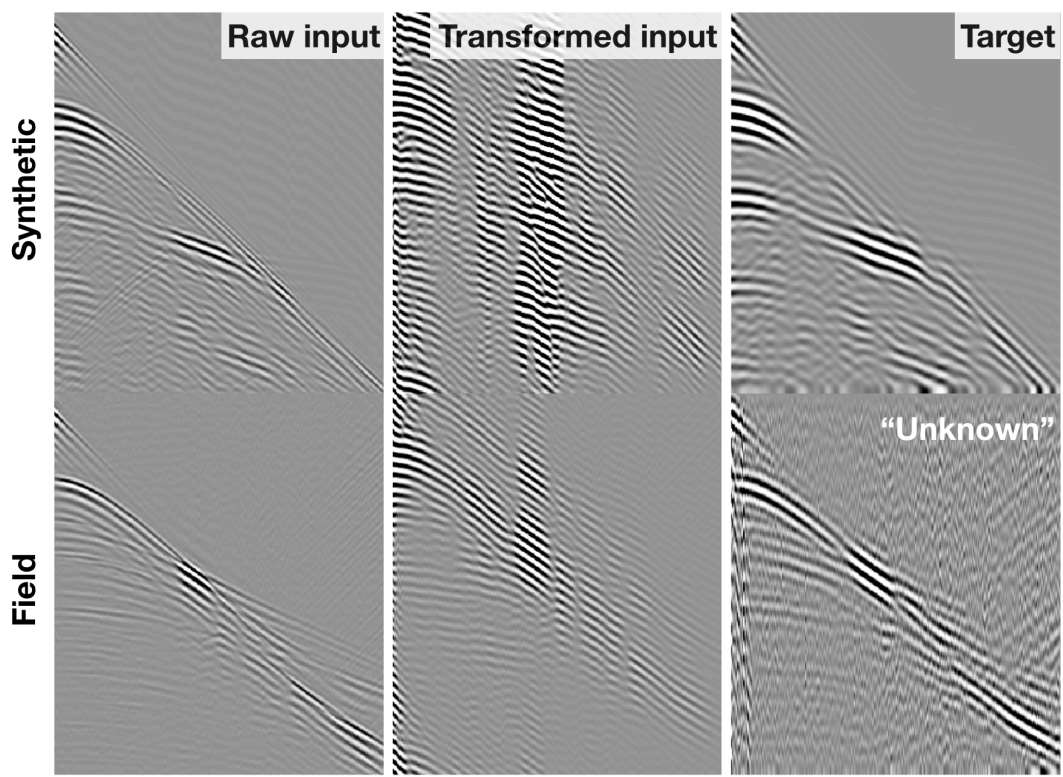
\multiplot{1}{input_target}{width=\columnwidth}{The original input and target data from synthetic and field datasets (first and last columns; the input data transformed by the proposed MLReal transformations (central column).}


\subsection{Deep learning framework}

We approach the frequency bandwidth extrapolation for the entire shot gathers as an image-to-image translation task. The design of a neural network should account for the need to capture long-wavelength patterns in the data since these are the data-domain projection of low-frequencies. For this reason, we follow \cite{wang2018image} to build a network that extracts multi-scale representations from the input volume by using dilated convolutions. A similar note about the need for wide-span convolutional kernels was made by \cite{sun2019extrapolated}. In particular,  the proposed architecture (Figure~\ref{fig:arch_low}) includes a three-column encoder featuring dilated convolutions with kernel sizes of 3, 5, and 7, followed by a convolutional decoder. In total, there are around 900k trainable parameters.
 
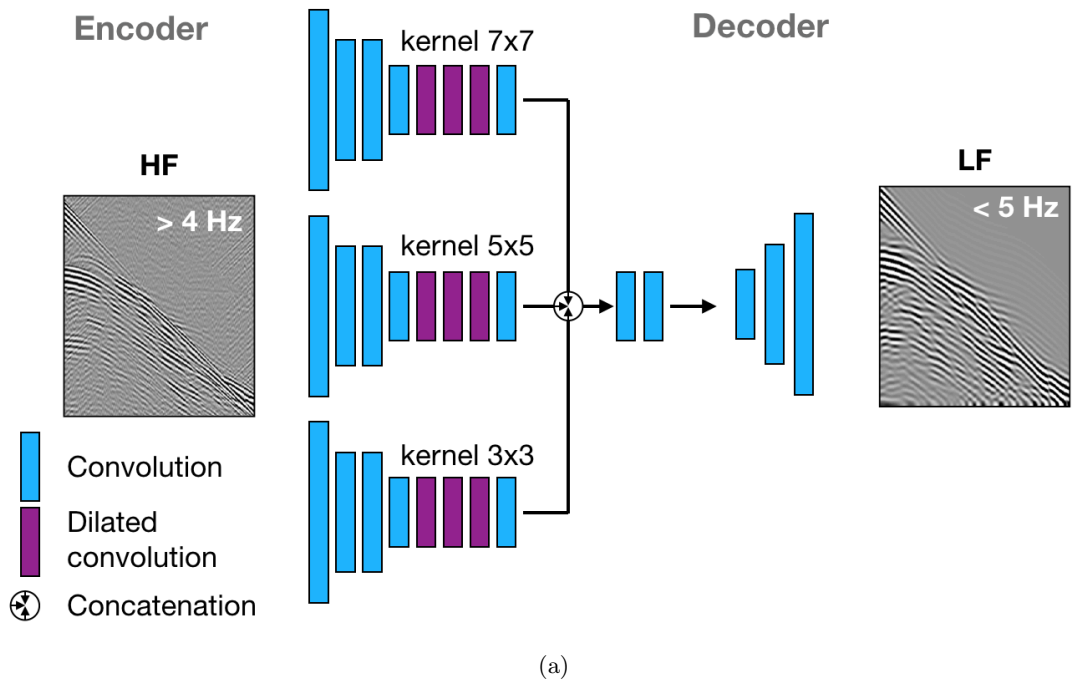
\multiplot{1}{arch_low}{width=\columnwidth}{The multi-column architecture for low-frequency extrapolation neural network model. This translates the high-frequency data, HF, into its low-frequency counterpart, LF.}

The training strategy implements the super-convergence concept by scheduling the learning rate according to \cite{smith2019super}, with the minimum and maximum learning rates of $1e-5$ and $1e-3$, respectively. We also find it helpful to initialize the network following \cite{7780459}, as well as to average predictions within an ensemble \citep{chollet2017deep} of 5 network initializations to reduce noise in the output data. One more note about training includes using the batch size of 4 since smaller batch sizes are prone to lead the inversion to non-sharp local minima \citep{keskar2016large}.

\subsection{Results}

Each shot gather used as input and output of the network is a single-channel image of $324 \times 376$, measuring the number of receivers in the marine streamer and the number of time samples, respectively. Receivers are spaced 25~m apart while the temporal sampling is coarsened to 8~ms. We create the training dataset of 3072 shot gathers by modeling 3 shots in each of 1024 random subsurface realizations. These shots are then split into partitions of 2765,  154, and 153 samples for training, validation, and testing, respectively. The set of random subsurface models is derived by distorting the layered models using elastic transforms, similar to \cite{kazei2021mapping}. The optimization by Adam \citep{kingma2014adam} stagnates after 80 epochs, delivering sufficient coveragence of the proposed training.

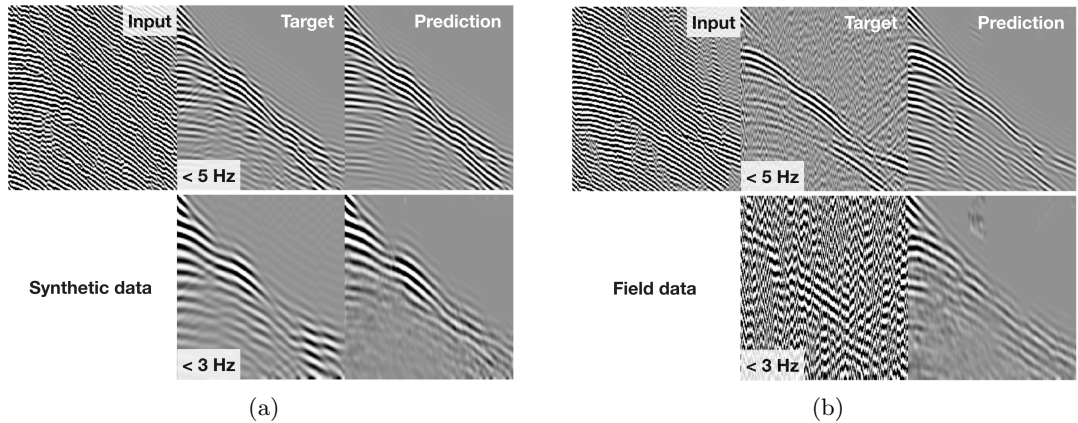
\multiplot{2}{syn_data,field_data}{width=0.45\textwidth}{Prediction results for synthetic (a) and field data samples (b). The top row contains the input data processed by the proposed transformations and the corresponding predictions of data below 5~Hz. The bottom row contains the same predicted data low-passed below 3~Hz.}

 The predicted low-frequencies $<$ 5~Hz indicate a good match for both synthetic (Figure~\ref{fig:syn_data}) and field datasets (Figure~\ref{fig:field_data}). This is expected due to the intentional overlap from 4 to 5~Hz between input and target frequency bands used to recover the amplitude of the signal when needed. We also low-pass the predicted data below 3~Hz to explore the regression capability of the trained network in the part of frequency spectra where data were not present. For the synthetic test, while weak reflections in the low-passed data are missing in the predictions, the general shape of the wavelet, as well as the early arrivals, are fairly well reconstructed. On the other hand, since real data suffers from a low signal to noise ratio at low frequency, like below 3 Hz, the predictions admitted higher signal-to-noise ratio, and reasonable smooth wavefields that can ultimately benefit waveform inversion. The high signal-to-noise ratio is courtesy of the synthetic data training where the output is free of noise and includes very low frequency. 
 
 In this example, we mainly conduct a proof-of-concept study of a geophysical application of our domain adaptation approach given by MLReal transformations (equations~\ref{eq:eq2} and~\ref{eq:eq3}), rather than seek the state-of-the-art bandwidth extrapolation accuracy. Realistic inference results by the same trained network on synthetic and field datasets suggest the viability of the approach for supervised regression tasks where only input data from the field dataset is available.

\section{Discussions}

The reason for the crosscorrelation with a reference trace is to balance the contributions from synthetic and real data to the input data to the network, without affecting the general features of the input data. In the microseismic example, this
operation also can help reduce the input data size. The fact that the input-to-the-network is effectively different from the original data is not an issue for NN models, as they adapt to any data we deem as input. What matters is whether the
input is consistent between the training (source) and application (target) data. Actually, the cross-correlation operation can enhance the data with features (from the cross-talk between unrelated events) that will further enrich the training data with information that can help in identifying the corresponding labels. For example, if a shot gather includes two reflections, the crosscorrelation of that shot gather
with a reference trace from it will result in two additional events from the crosstalk of the two events, and these added events will be different from
one shot gather to another depending on the distance between the two events and their moveouts. Such additional events will add to the information embedded in the input data
that the NN model can use to learn to identify the corresponding labels.

However, the application of the MLReal transforms in equations~\ref{eq:eq2} and~\ref{eq:eq3}, and specifically, the crosscorrelation with a reference trace will
alter the vertical axis of the input data, moving the energy closer to zero lag, or in other words, retaining only information on the relative location of events in the vertical
dimension (whether the vertical dimension is time or depth), not their absolute locations. If the task is a classification of features not related to the absolute value of the vertical axis, like in the microseismic source location task
we shared, this process will not affect the ability of trained models
to classify the real data, since the real data are exposed to the same operations. This holds for any task that does not depend on the actual scale of the vertical axis (time or depth) and relies more on the relative location of events vertically, and on lateral behavior of features. This limitation should not be a critical obstacle for seismic data recorded on the
Earth's surface as our resolution of depth (the vertical axis) has always been limited. This limitation often manifests itself in the misties we face between seismic and well depths. In other words, the vertical scale, in many seismic tasks, is poorly represented in the data.

Nevertheless, there is a relatively straightforward remedy for such a limitation that would allow us to apply the proposed approach to a wider range of tasks. If the task involves the same size data (in the vertical dimension) for the input and output to the network,
like in denoising, super-resolution, and even interpolation of missing traces, the crosscorrelation operation can be applied to both the input and output data. The reference trace should be taken from the same channel location from the input data (as the input data are available for training and application). For prediction, in this case, we will need to apply an inverse crosscorrelation on the predicted data using the same location
reference trace used in the training, but taken from the input to the prediction. The inverse crosscorrelation will transform the output of the prediction back to the
form of the section we desire. Figure~\ref{fig:diagramOuput} outlines these steps. The correlation and inverse correlation steps could be done in the Fourier
domain, as the inverse correlation, in this case, implies dividing each output trace by the reference trace. Kinematically, this implies that in the training, we subtract the phase of the reference trace from the input data (crosscorrelation) and then we add it back again to the output of the NN model in the inference stage (inverse crosscorrelation). 
If the amplitudes of events are not critical, the inverse crosscorrelation can be replaced with convolution, which has exactly the same effect on the phase.

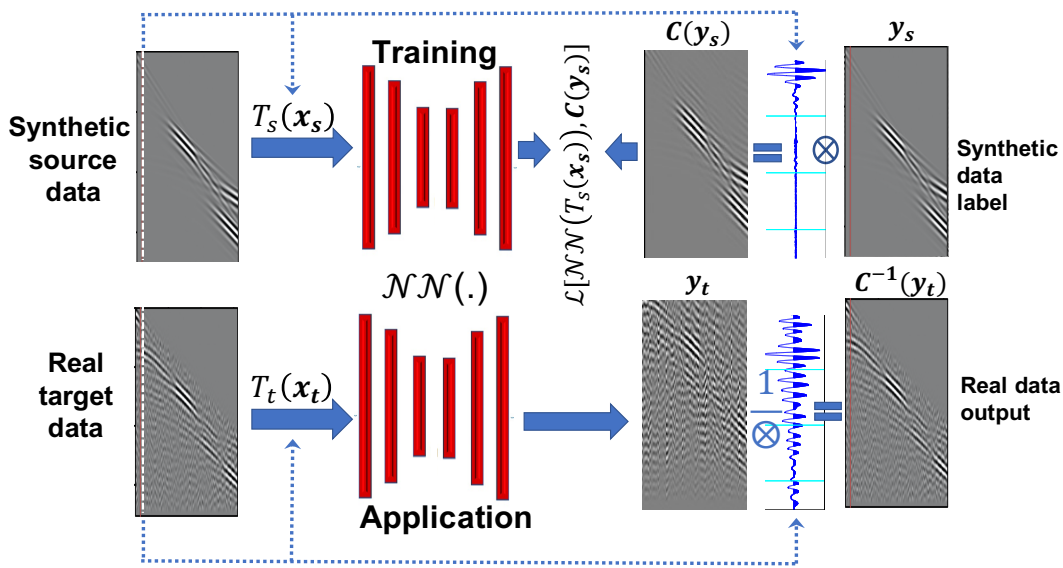
\plot{diagramOuput}{width=0.95\textwidth}{A diagram describing the steps applied to the output of the NN network to allow the proposed approach to work for applications like denoising and missing trace recovery when the training is performed on synthetic data. For training (top row), the original label ${\bf y_s}$ is correlated with a reference trace from a fixed location from the
input data, as the dashed line indicates, to provide the output for the training $C({\bf y_s})$. This same reference trace is used in the transform $T_s$ given by equation~\ref{eq:eq2}. For inference (bottom row), the output of the NN model,${\bf y_t}$, is inversely correlated with a fixed location reference trace from the input, as the dashed line indicates,
to provide the final output.}

Finally, there is nothing in the proposed MLReal transformations that can prevent their 
applications to waveforms (seismic or electromagnetic) at any scale.
The proposed transformations are applied to the vertical dimension of a 
multi dimensional input to the network, and in the shared examples the vertical dimension was the time axis. The microseismic example can be considered as a miniature test
of a potential Earthquake location task. So if the objective is to study and 
monitor earthquake locations in a certain region using a set of seismic stations \cite[]{10.1093/gji/ggy261,10.1093/gji/ggx472},
we can use a global model \cite[]{10.1093/gji/ggaa253} to simulate synthetic data at the seismic stations, albeit elastic (multi component), from random sources (with random source moment tensors) at the monitored area of interest. In this case, one of the stations are picked as the reference station for crosscorrelation and fixed for both synthetic and real data. Then the MLReal transformations will migrate the real data features (events not reproduced by the global model) to the synthetic training set, and vice versa, as shown for the microseismic example in Figures~\ref{fig:field_data_convolve_with_mean_autocorrelation2}
and~\ref{fig:synthetic_data_convolve_with_mean_autocorrelation2} .

\section{Conclusions}

We proposed a novel technique to precondition the synthetic training data set for a supervised neural network optimization so that the trained model works better on
real data. The concept is based on incorporating as much information from the real data into the training without harming the synthetic
data features crucial for the prediction. Considering the two data domains (synthetic and real), we specifically cross-correlate an input section from one domain of data with a reference trace from that data followed by convolution with an
autocorrelated section from the other domain. For training the NN model, the input section is from the synthetic data domain, and for the application (inference) of the NN model, the input section is from the real data domain. A test of this approach on a microseismic source location task using input waveforms helped us improve the application of the NN model on real data. Another application, in which we train an NN to predict low frequencies, the preconditioning helped improve the prediction on real data.

 

\section{Acknowledgments}
\vspace{-0.15in}
We thank Umair bin Waheed from KFUPM, Frantisek Stanek from Seismik, and Claire Birnie and Yuanyuan Li from KAUST for helpful discussions. 
We thank Microseismic, Inc. and Newfield Exploration Mid-Continent, Inc. for graciously supplying the data. We thank CGG for the marine dataset.
We also thank KAUST for its support and the seismic wave analysis group (SWAG) for constructive discussions.


\bibliographystyle{seg}
\bibliography{ms}

\end{document}